PREPRINT

# Extreme Ultraviolet (EUV) Sources for Lithography based on Synchrotron Radiation

Guiseppe Dattoli[1], Andrea Doria[1], Gian Piero Gallerano[1], Luca Giannessi[1], Klaus Hesch[2], Herbert O. Moser[6], Pier Luigi Ottaviani[1], Eric Pellegrin[3], Robert Rossmanith[2*], Ralph Steininger[2], Volker Saile[4], Jürgen Wüst[5]

[1]ENEA INN-FIS-LAC, Frascati, Italy

[2]Synchrotron Radiation Research Group, [3] Institute of Solid State Physics, [4] Institute of Microstructure Technology, [5] Technology Transfer and Marketing, Forschungszentrum Karlsruhe

[6] Singapore Synchrotron Light Source SSLS, National University of Singapore

[*]Corresponding author, Forschungszentrum Karlsruhe, Synchrotron Radiation Research Group FGS, P. O. Box 3640, D-76021 Karlsruhe, Germany, Tel. ++49 7247 82 6179, Fax ++49 7247 82 6172, e-mail rossmanith@anka.fzk.de

Submitted to Nuclear Instruments and Methods A

---

Work supported by the German Ministry for Research and Education BMB+F under contract No. 01 M 3103 A

**Abstract:**

The study presented here was initiated by a discussion to investigate the possibility of using synchrotron radiation as a source for the Next Generation Lithography (NGL) based on the EUV-concept (Extreme Ultra-Violet; here 13.5 nm or 11.3 nm radiation, respectively). The requirements are: 50 W, 2% bandwidth and minimal power outside this bandwidth. Three options were investigated. The first two deal with radiation from bending magnets and undulators. The results confirm the earlier work by Oxfords Instrument and others that these light-sources lack in-band power while emitting excessive out-of-band radiation. The third approach is a FEL (Free Electron Laser) driven by a 500 MeV linear accelerator with a superconducting mini-undulator as radiation emitting device. Such a device would produce in-band EUV-power in excess of 50 W with negligible out-of-band power.

---

1. **Introduction**

Lithography, the technique for manufacturing microelectronic semiconductor devices such as processors or memory chips, presently uses deep UV (DUV) radiation. The main radiation source is the 193 nm line of an ArF excimer laser [1]. Future sources will be $F_2$ lasers at a wavelength of 157 nm and eventually $H_2$ lasers at a wavelength of 127 nm.

In addition, advanced lithography technologies (Next Generation of Lithography: NGL) based on EUV, X-ray photons, electrons, and ions are being investigated by chip makers and equipment manufacturers.

The competing technologies are: SCALPEL electron lithography (Scattering with Angular Limitation in Projection Electron-Beam Lithography) [2], Ion Projection Lithography [3], X-ray Proximity Lithography [4] and Extreme UV Lithography [5]. The latter is being considered as one of the most promising. In the US a program to develop this technology was set up as early as in 1994 by the EUV LLC (Limited Liability Corporation) in cooperation with the VNL (Virtual National Lab). Members of VNL are LLNL (Lawrence Livermore National Lab), LBNL (Lawrence Berkeley National Lab) and Sandia National Lab. In Japan the ASET consortium was funded (Association of Super-Advanced Electronics Technologies) [6].

During the research phase the needs for a EUV source suitable for future production lines were identified. The main requirements are:

Wavelength: 13.5 nm (=92 eV) or 11.3 nm

Bandwidth: 2%

Output power: 25 W (first step) and later 50 W

In addition, the power radiated outside this band has to be less than 500 W to avoid thermal problems on the optics. The development of a suitable source is one of the big challenges in EUV lithography.

Basically, powerful sources of EUV photons may be based on either plasmas [7] (produced by laser irradiation of matter or by gas discharges) or on relativistic electrons (synchrotron radiation).

In Europe, the development of synchrotron radiation-based EUV sources [8] was partly supported by the European Union within the framework of the EUCLIDES program [9]. Similar investigations were performed in Japan [10] and the USA [11]. The studies showed that conventional storage rings with and without additional magnets (normal conducting or superconductive wigglers or undulators) do not fulfil all the above-mentioned specifications for the EUV source.

The German Federal Ministry of Education and Research initiated at the beginning of 2000 a program on plasmas generated by lasers or gas discharges as sources of EUV light for the next generation lithography (NGL). In the initial phase of this project it was felt that sources based on synchrotron radiation should be reconsidered. In the first quarter of 2000 the authors presented their report. The present paper is a shortened version of this report. The result was that among all possible sources based on synchrotron radiation only a Free Electron Laser can meet the above mentioned stringent requirements at 13.5 nm. In summer 2000 a group at DESY published independently a paper in which the design of a SASE Free Electron Laser source for lithography at 70 nm is described [33 ] confirming at least in principle the viability of the FEL concept.

## 2. Incoherent radiation from the storage rings

In the following the results already obtained in the EUCLIDES study are summarized for reference. A model storage ring is shown in fig.1. The parameters which are needed to calculate the emitted photon intensity at 13.5 nm within the required bandwidth of 2% are

- electron energy
- magnetic field strength
- electron current

The maximum storable current depends on two limitations: beam instabilities and intrabeam scattering (Touschek-Effect) [12]. Beam instabilities can be defeated by feedback systems. The Touschek lifetime for an unpolarized beam is approximately

$$\frac{1}{\tau[\mathrm{sec}]} = \frac{\sqrt{\pi} r_e^2 c N C(\zeta)}{\sigma'_x \gamma^3 \varepsilon_{acc}^2 V} \qquad (1)$$

where $r_e = 2{,}8.10^{-15}$ m (classical electron radius). N is the number of particles per bunch. Assuming a 500 MHz RF frequency N is equal to $1{,}25.10^{10}$ for a stored beam of 1 A. $\gamma$ is the ratio between energy and rest energy.

The rest of the parameters describes the particle density in relation to the region in which the particles are stable, the so-called energy acceptance.

$\varepsilon_{acc}$ is the energy acceptance of the storage ring. A particle gets lost when the scattering is so violent that a particle changes its energy by more than that. The scattering probability depends on the density of the particles in the bunch. The parameter $\zeta = (\varepsilon_{acc}/\gamma\sigma'_x)^2$. $\sigma'_x$ is the divergence in the beam.

For $\zeta \leq 10^{-2}$ the following approximation is valid $C(\zeta) \cong -\ln(1{,}732\zeta)-1.5$. The bunch volume V is $8\,\pi^{3/2}\sigma_x\sigma_y\sigma_L$. Typical Touschek life times for a 1 A beam are summarized in Table I.

The strong dependence of the Touschek effect on the energy indicates that the preferred storage rings are operating at higher energies: 0.3 GeV and higher.

The spectral power ΔP of the emitted synchrotron radiation in Watts per eV, per mrad horizontal angle $\vartheta$ and integrated over the vertical angle is given by formula (2) [12], [13].

$$\frac{\Delta P}{\Delta \vartheta}\left[Watt / mrad\vartheta / eV\right] = 8.73 \frac{E^4[GeV] I[A]}{r[m]} G_2(y) \qquad (2)$$

with $G_2(y) = y^2 \int_y^\infty K_{5/3}(\eta) d\eta$ and $y = E_{Phot} / E_c$

I is the stored beam current, E is the energy of the stored beam and K is the modified Bessel function. $E_{Phot}$ is the photon energy and $E_c$ is the so-called critical photon energy

$$E_c\,[eV] = 2218.3 \; \frac{E^3[GeV]}{r[m]}$$

r is the bending radius. r and the bending field B in Tesla are related by the equation:

$$r[m] = 3.34 \frac{E[GeV]}{B[T]} \qquad (3)$$

The power emitted at 13.5 nm per mrad horizontal angle within a 2% bandwidth is according to (2)

$$\frac{\Delta P}{\Delta \vartheta}\left[Watt / mrad\vartheta\right] = 0.1746 \frac{E^4[GeV] I[A]}{r[m]} G_2(y) \qquad (4)$$

Figs. 2 and 3 show the results of equation (4). The assumed current is 1 A in all cases.

In conclusion it can be said that for ca. 100 eV photons the spectral density has a maximum at fields near 1.5 to 2 T. It follows from fig. 3 that the spectral power increases the higher the energy is. The optimum values can be reached with room temperature magnets (1.5 T) and high energies (in other words with fairly large machines).

The maximum angle over which photons can be collected is 6.28 rad (the full circumference of the storage ring). The maximum in-band power as a function of energy and field strength is shown in fig. 4 for a stored beam of 1A. Fig. 5 shows 2 D cuts of figure 4.

From these curves it is obvious that the 50 W requirement with a stored beam of 1 A can only be met at energies significantly above 1 GeV. The maximum collectible power at an energy of 0.6 GeV is 27 W. This in-band power has to be compared with the total radiated power:

$$P_T[kW] = 88.5 \frac{E^4[GeV]I[A]}{r[m]} \tag{5}$$

which is for 0.6 GeV and 1.5 T (r = 1.336 m) ca. 8.6 kW. The power ratio (defined as in-band power P/total power $P_T$) is

$$\frac{P}{P_T} = 0.01239.G_2(y) \tag{6}$$

The power ratio is shown in fig. 6. The maximum values are obtained at low fields. This argument confirms that high beam energy and low magnetic fields are the optimum parameters. A storage ring of 0.6 GeV and a field of 1.5 T might be a fair compromise to obtain a total in-band power of more than 25 W.

It is obvious all the formulas mentioned in the previous chapter valid for bending magnets are also valid for wigglers. The wiggler has two advantages over a bending magnet. Firstly, the photons are emitted into a cone centred around the direction of motion of the beam. The collection of photons is easier with a wiggler than with a bending magnet. Secondly, since there is no net deflection, it is easier to choose the optimum field.

Wigglers with a small maximum beam deflection angle α are called undulators. The K-value is defined in the following way:

$$K = \alpha.\gamma = 0.94.\ B[T].\ \lambda_u\ [cm] \qquad (7)$$

Constructive interference in the vicinity of the beam axis happens when:

$$\lambda_{Phot} = \frac{\lambda_u}{2n\gamma^2}\left(1 + K^2/2 + \gamma^2\theta^2\right) \qquad (8)$$

$\theta$ is the angle between electron beam axis and the photon detector.

A measured spectrum of the first harmonics of an undulator depending on the angle is shown in fig. 7. In fig. 7 angle and photon energy are clearly related (depending on the emittance of the beam). This is described by (8) [28].

The undulator condition (8) has to be fulfilled for 13.5 nm. This condition limits the number of possible solutions for the period length. In addition, a general rule states that the period length should not be shorter than 4 times the gap width of the undulator. If this rule is not observed, than the field acting on the beam becomes too small [14].

The formula used in Table I for the total power radiated from an undulator is

$$P[W] = \frac{7.26E^2[GeV]I[A]N_u K^2}{\lambda_u[cm]} \qquad (9)$$

$N_u$ is the number of periods. The calculated in-band power for an undulator is shown (as an example) in fig. 8. The parameters of different undulators are summarized in Table II. The first harmonics of all undulators is close to 13.5 nm.

Despite the fact that the maximum obtainable power does not fulfil the stringent requirement the undulator has clear advantages over wigglers.

The K-values in Table I are in the order of 1 to 2. According to formula (7) the magnetic fields of the undulator are larger than 1 T. These values are larger than those achieved with conventional permanent magnet undulators. In Brookhaven [15] and Karlsruhe [16] independent concepts for using superconductors rather than arrays of permanent magnets have been under discussion. Recently Karlsruhe together with a group at Mainz [20] have been able to demonstrate the viability of such a concept under normal beam conditions. Fig. 7 shows the measured spectrum from these experiments.

Fig. 9 shows the principle of a superconductive undulator. The field is generated by a superconductive wire in an iron matrix (darker parts in fig. 9). The superconductive wires are close to the beam. The undulator is indirectly cooled by liquid helium not shown in figure 9. The parameters for this specific undulator are: period length 14 mm and K=2 (1.5 T) [30]. The calculated undulator field is shown in fig. 10.

### 3. The Free Electron Laser approach

It has been shown in the previous chapters that the specifications for the source defined in the introduction, 50 W within a 2 % bandwidth at 13.5 nm, is barely achievable to obtain with a conventional synchrotron radiation source.

In 1951 Motz was the first to point out that the intensity of a photon beam emitted by electrons can be increased by coherent superposition [18]. The logic is as follows. If each electron emits a photon the resulting electric field $E_{total}$ is:

$$E_{total} = \sum_n E_n \tag{10}$$

$E_n$ is the electric field of the individual photons. The intensity is proportional to $E^2_{total}$. When the phases of the photons have a random distribution (incoherent light) the cross terms cancel and the averaged sum is

$$I = \left\langle \sum_i E_i . \sum_j E_j \right\rangle = N.E^2 \tag{11}$$

where N is the number of electrons.

When the phases of the electrons are identically and they are not randomly distributed the cross-terms do not disappear and the intensity is $N^2$ times the intensity of a single electron. This is obviously the case when the electrons are concentrated in bunches. The length of these bunches (so-called micro-bunches) must be smaller than the wavelength. The micro-bunches are separated by a multiple of a wavelength.

If this argument is turned around, then most of the intensity of a conventional synchrotron radiation source is destroyed by incoherence or, if expressed in other terms, by the random distances of the emitting electrons. When the electrons have distances which are smaller than the emitted light wave the intensity can be increased by an enormous factor (N is a very big number).

Since 1951 this principle has been experimentally investigated with great success by various groups and this has changed dramatically the design of synchrotron light sources [19], [20] and beam diagnostics tools [21].

In order to operate a FEL effectively the following conditions on the emittance have to be fulfilled [29].

a.) Particles with an angle to the beam axis do not fulfil the resonance condition (8). The electron velocity in the direction of the axis is changed by $-x'^2/2$ ( $x'$ is the angle relative to the axis). In order to keep the electron within a half-wave over the whole undulator length, the condition

$$x' \leq \left( \frac{\lambda_{Phot}}{2L} \right)^{1/2} \tag{12}$$

has to be fulfilled, where L is the length of the undulator [31] .

b.) The spot-size of the optical mode is given by

$$x \leq \left( \frac{\lambda_{Phot} R}{\pi} \right)^{1/2} \tag{13}$$

where R is the Raleigh length (the distance in which the area of a diffracted wave doubles). Typically the Raleigh length is one-half of the interaction length L.

The restrictions c.) and d.) are usually combined to one requirement [22]

$$\varepsilon \le \frac{\lambda_{Phot}}{4\pi} \tag{14}$$

Equation (14) requires that the horizontal and vertical emittance of the beam has to be smaller than 1.07 nm. In a linac the emittance shrinks with energy (adiabatic damping)

$$\varepsilon = \varepsilon_n / \gamma \tag{15}$$

where $\varepsilon_n$ is the so-called normalized emittance. The magnitude of the normalized emittance depends on the gun. For a photo-cathode gun $\varepsilon_n$ is close to $10^{-6}$ m.rad (depending on current, bunch length etc.) [23]. Following equation (15) $\gamma$ has to be 1000 or higher (linac energy equal or above 500 MeV). Assuming a gradient of 20 MeV/m, the linac is 27.5 m (or close to 30 m) long. For 0.52 GeV and K = 1.4 the period length of the undulator is 1.43 cm according to equation (8). The peak field is circa 1.05 T according to equation (7). Linacs with more than 40 MeV/m are available, so that the minimum length of the linac is about 15 m.

The SASE FEL [32] is generally described by analytical methods. The fundamental parameter in this description is the gain length $L_G$, the length in which the FEL power increases by a factor of e. The gain length depends on the power density of the emitted light. The power density is a function of the undulator properties (K-value, period length), the beam properties (peak current, energy, $\beta$-functions, emittance, energy spread etc.) and the properties of the optical beam (diffraction). In order to separate the different influences the following parameters are introduced:

$$L_G = \frac{\lambda_u}{4\pi\sqrt{3}\rho\chi S} = \frac{L_{G0}}{\chi S} \tag{16}$$

where $\rho$ is the so called Pierce parameter

$$\rho = \frac{1}{4\pi}\left[\frac{2\pi^2}{\gamma^3}(JJ[1,K].\lambda_u.K)^2 \frac{I_{peak}}{I_A}\frac{1}{2\pi\beta\varepsilon}\right]^{1/3} \tag{17}$$

The Pierce parameter describes the emission of synchrotron radiation. $I_A$ is 17 kA (Alven current) and the rest of the parameters are explained in previous equations. $\chi$ and S are correction parameters describing the influence of the diffraction and the energy spread.

The diffraction effects are described by the parameter S in (16):

$$1/S = \left[1 + \frac{L_{G0}^2}{\beta^2}\left[\frac{\lambda_{Phot}}{4\pi\varepsilon}\right]^2\right] \tag{18}$$

and produce the curve shown in fig. 11 [24]. The curve has a minimum close to a beta of 0.3 m with a slight slope towards higher beta-functions.

A beta function of 2 m throughout the whole undulator is an acceptable compromise. The undulator shown in fig. 9 has to be modified in such a way that it focuses in both directions. The focusing in both directions in an undulator with permanent magnets was demonstrated at DESY for the first time [35]. A similar effect can be achieved for the superconductive undulator either by shaping the iron poles in an appropriate way. A study on SASE and superconductive undulators can be found in [26].

Up to now the influence of the energy spread was not taken into account. The energy spread leads to a broadening of the laser line and finally to a loss of gain. The influence of the energy spread is described by the function $\chi$ in a complex way. $\chi$ is determined by three parameters:

$$\mu_\varepsilon(\beta) = \frac{2\sigma_\varepsilon}{\rho(\beta)}$$

$$\mu(\beta) = \sqrt{3}\frac{L_{G0}(\beta)}{\lambda_0}\frac{\pi^2 K^2 \varepsilon_n \beta}{\gamma^2 \lambda_u^2}$$

$$\mu_1(\beta) = \sqrt{3}\frac{L_{G0}(\beta)}{\lambda_0}\frac{\varepsilon_n}{\gamma\beta} \tag{19}$$

All 3 parameters depend on β and on the energy spread $\sigma_\varepsilon$. The parameters have to fulfil an integral equation and a solution is only possible by numerical techniques. The function $\chi(\beta)$ is a solution of the integral equation:

$$\left| \int_0^\infty e^{-\frac{\sqrt{3}+i}{2}\chi s} e^{-\frac{(\pi\mu_\varepsilon s)^2}{2}} \frac{1}{(1-i\pi\mu s)} \frac{1}{(1-i\pi\mu_1 s)} ds \right| = \chi \tag{20}$$

$\chi$ depends somewhat on $\beta$ but strongly on the energy spread.

The FEL process starts from the radiation emitted in the first gain length of the undulator. The number N of undulator periods in the first gain length (equation (16)) is

$$N_G = \frac{1}{4\pi\sqrt{3}\rho} \tag{21}$$

and the spontaneous peak power P emitted during the first gain length (peak current $I_{Peak}$):

$$P(\beta)[W] = 1.48x10^{19} E_{electron}^2[GeV] N_G \frac{JJ[1,K]^2 \beta\varepsilon_n I_{Peak}}{L_G^2 \lambda_{Phot}} \tag{22}$$

For a relative energy spread of $10^{-4}$ this value is circa 19 W for a beta-function of 2.5 m.

The development of the power along the undulator axis z is described by

$$P(z) = \frac{P(\beta)}{9} e^{z/L_G} \tag{23}$$

The amplification is stopped by an undulator with constant period length at the saturation length $z_{sat}$

$$z_{sat} = L_G \ln\left(9\frac{P_{peak}}{P(\beta)}\right) + L_G \tag{24}$$

with

$$P_{Peak} = 10^9 \rho I_{Peak}[A]E[GeV]$$

In the following it is assumed that the beta-function is 0.5 m. As shown before, all results depend strongly on the beta function.

The development of the peak power, including saturation, is

$$P_z = \frac{\frac{P(\beta)}{9} e^{\frac{z}{L_G}}}{1 + \frac{P(\beta)}{9P_{Peak}}\left(e^{\frac{z}{L_G}} - 1\right)} \tag{25}$$

The dependence of the peak power on the undulator length is shown for two cases (energy spread of $10^{-4}$ and an energy spread of $5.10^{-4}$) in figs. 12 and 13. The peak current $I_{peak}$ is 200 A. Obviously, the final peak power is the same in both cases. The energy spread only defines the length of the undulator [27].

The peak power of one pulse is ca. $1.33.\ 10^8$ W. One pulse with an assumed bunch length of 3 psec produces an energy of $\sqrt{2\pi}\ .3.10^{-12}\ 1.33\ 10^8$ J or about 1 mJ. In order to produce a cw power of 50 W, a pulse repetition rate of 50 kHz is required.

The required average current is fairly modest. Working on the basis of a 1.5 GHz linac RF system (bucket repetition of time of 0.67 nsec) the average current is

$$200(3/667)(5.10^4/1,5.10^9)\ \text{A} = 30\mu\text{A}.$$

## 4. Possible Layout of the FEL source

A possible layout of the EUV laser system for a wafer fab is shown in fig. 14. It is assumed that the EUV source (linac, undulator etc.) will be located in the basement of the factory. The EUV radiation enters the clean room via evacuated pipes which come up through the floor. The normalized emittance of the beam (assumed to be$10^{-6}$) determines the energy of the linac: 500 MeV.

The accelerating structures can be either superconductive (Nb-cavities) or normal conducting (Cu-cavities).

Normal conductive cavities allow simple and short structures: energy gains of up 40 MeV/m and higher are possible. The length of the linac would be less than 15 m.

The accelerating gradient for superconductive linacs is at the moment ≥20 MeV/m and, as a result, a superconducting linac will be almost twice as long as a room temperature linac.

The accelerated beam is directly sent to a 11m long SASE undulator. All bends along the trajectory have to be isochronous in order to prevent bunch lengthening.

The installation of most of the equipment in auxiliary and/or distant rooms, such as the basement is an integral part of the following layout considerations. Fig. 14 shows one possible way of distributing the EUV light. A central linac provides a distributed undulator system with an electron beam. The beam is switched by magnets to the undulators. The fact that several SASE superconducting undulators are fed from one linac reduces the capital cost per stepper.

## Conclusion

The German Federal Ministry of Education and Research initiated a program on plasmas generated by lasers or gas discharges as sources of EUV light (50 W at 13.5 nm, bandwidth 2%) for the next generation lithography (NGL). In the initial phase of this project it was felt that sources based on synchrotron radiation should be reconsidered. The aim of this report is to investigate such sources.

The report starts with investigations into the emitted power of small storage rings with energies of less than 0.6 GeV. The total emitted power collected over the entire circumference is less than 50 W (stored current of 1 A).

In a next step storage rings were equipped with wigglers. It is easier to collect the wiggler radiation but the conclusions are similar: the total emitted power is insufficient.

In the following step storage rings with undulators have been studied. Under certain circumstances these devices have clear advantages over the wiggler system. In undulators the emitted photons can interfere coherently. This fact makes it possible to amplify the intensity within the required bandwidth and minimize it outside. In order to optimise the output power, superconducting mini-undulators are required.

Free Electron Lasers (FELs) consisting of linacs and undulators produce light with a high degree of coherence and of high power. Unwanted out-of-band-radiation is almost completely eliminated. The study shows that the so-called SASE technique (Self Amplified Stimulated Emission) can easily produce the required EUV power. The SASE effect was already observed experimentally at wavelengths as low as 80 nm.

As a result, synchrotron radiation (mainly FELs) can easily fulfil the stringent requirements for the Next Generation of Lithography based on EUV when suficet space is forseen in a wafer fab.

**Acknowledgements**

This study is based on numerous discussions with and contributions from many colleagues. The authors would like to thank them. It is impossible to mention all names of the individuals who contributed to this study. Our special thanks go to DESY (Prof. Schneider, Prof. Materlik, Dr. Rossbach, Dr. Pflueger and the SASE FEL team), to ESRF (Dr. Elleaume and his team), BNL (Dr. Ben-Zvi and colleagues), Swiss Light Source (Prof. Wrulich, Dr. Ingold), ENEA (Prof. Renieri), Elettra (Dr. Walker and colleagues), JLab (Dr. Neil), University of Virginia (Prof. Norum, Prof. Gallagher), LBNL (Dr. Jackson, Dr. Robin and colleagues), Duke University (Profs. Edwards and Litivenko), UCLA (Prof. C. Pellegrini), ACCEL (Drs. Klein, Krischel, Schillo and Geisler), and many others.

| Energy [GeV] | $\gamma$ | $\zeta$ | $C(\zeta)$ | $\tau$ [sec] |
|---|---|---|---|---|
| 0.6 | 1174 | $1{,}85.10^{-4}$ | 6.54 | 24064 |
| 0.4 | 783 | $4{,}13.10^{-4}$ | 5.74 | 8117 |
| 0.3 | 587 | $7{,}35.10^{-4}$ | 5.17 | 3805 |
| 0.2 | 391 | $1{,}65.10^{-3}$ | 4.36 | 1334 |
| 0.1 | 196 | $6{,}00.10^{-3}$ | 3.01 | 246 |

Table I: Touschek lifetime for various beam energies. Beam current 1 A. $\varepsilon_{acc}$= $5.10^{-3}$, $\varepsilon_{hor}$=500nm, $\varepsilon_{vert}$ = 10 nm.rad, $\beta_{hor}$ = 5 m, $\beta_{vert}$ = 10 m, $\sigma_L$ =3 cm .

| Energy [GeV] | K | $\lambda_u$ [cm] | Total emitted power [W] | In-band power [W] | Power ratio |
|---|---|---|---|---|---|
| 0.3 | 1 | 0.62 | 105 | 2.5 | $2.4.10^{-2}$ |
| 0.4 | 1.5 | 0.78 | 335 | 3.9 | $1.2.10^{-2}$ |
| 0.4 | 1 | 1.1. | 106 | 2.5 | $2.4.10^{-2}$ |
| 0.6 | 2 | 1.24 | 843 | 4.3 | $5.1.10^{-3}$ |
| 0.6 | 1.5 | 1.49 | 544 | 3.9 | $7.2\ 10^{-3}$ |
| 0.6 | 1 | 2.48 | 105 | 2.5 | $2.4.10^{-2}$ |

Table II Characteristics of various 100 period long undulators. The current is 1 A in all cases.

Fig. 1: Model storage ring source

Fig. 2: Spectral power at constant field with the energy as parameter

Fig. 3: Spectral power at constant energy with the magnetic field as a parameter

Fig. 4: In-band power versus energy and magnetic field as in fig. 5 ( energy range 0.2 to 0.6 GeV)

Fig. 5: In-band power versus magnetic field with the energy as a paramete (2D cuts of fig .4)

Fig. 6: Power ratio as a function of energy and bending field strength

Fig. 7 Angle dependance of the measured X-ray spectrum of an undulator [16]

Fig. 8 Undulator: beam energy 0.6 GeV, K=2, $\lambda_u$ =1.24 cm. The maximum in-band output power isabout 4.8 Watt. Calculated with the program URGENT [17]

Fig. 9: Layout of the superconductive miniundulator (shown from two perspectives). The dark red material is iron, the lighter coloured material depicts superconductive wires. The beam travels in the gap between the two undulator poles. The current direction through the wires alters from wire to wire generating the undulator field.

Fig. 10: Undulator field (calculated)

Fig. 11: Influence of diffraction effects on the gain length

Fig. 12: Peak power versus undulator length for an energy spread of $1 \cdot 10^{-4}$

Fig. 13: Peak power versus undulator length for an energy spread of $5 \cdot 10^{-4}$

Fig 14: Chain-type layout of an FEL source

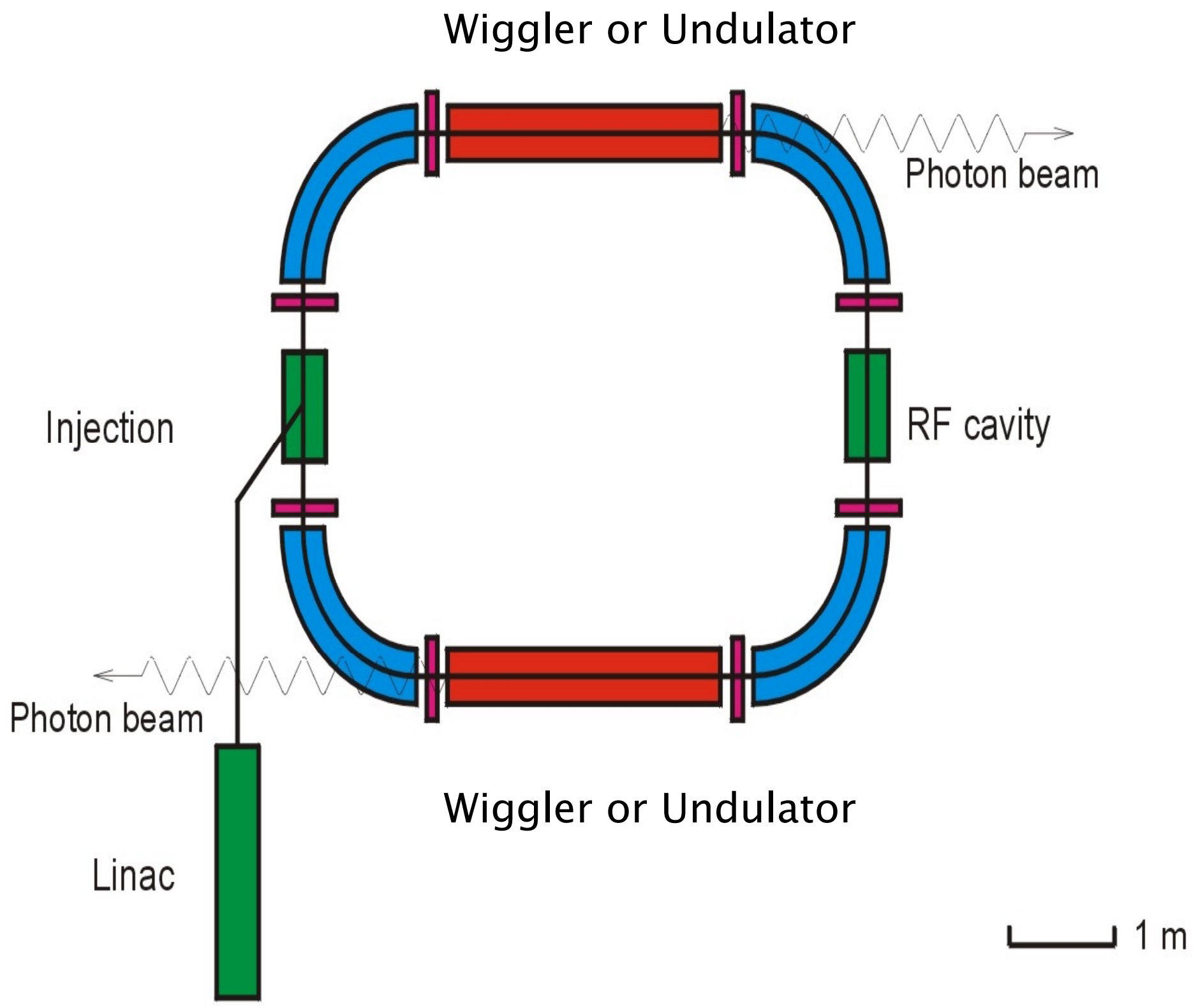

*Fig.1*

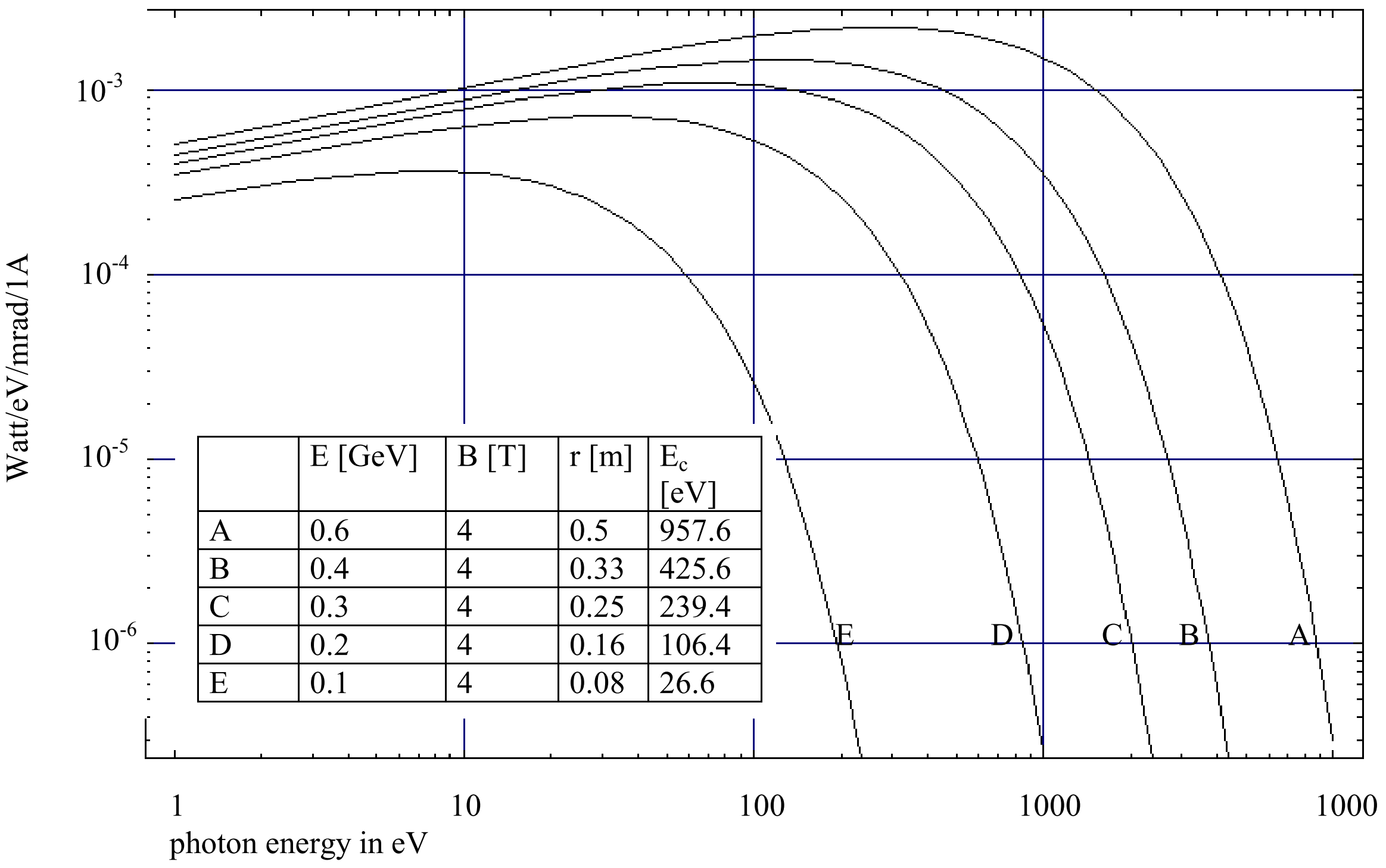

| | E [GeV] | B [T] | r [m] | $E_c$ [eV] |
|---|---|---|---|---|
| A | 0.6 | 4 | 0.5 | 957.6 |
| B | 0.4 | 4 | 0.33 | 425.6 |
| C | 0.3 | 4 | 0.25 | 239.4 |
| D | 0.2 | 4 | 0.16 | 106.4 |
| E | 0.1 | 4 | 0.08 | 26.6 |

*Fig. 2*

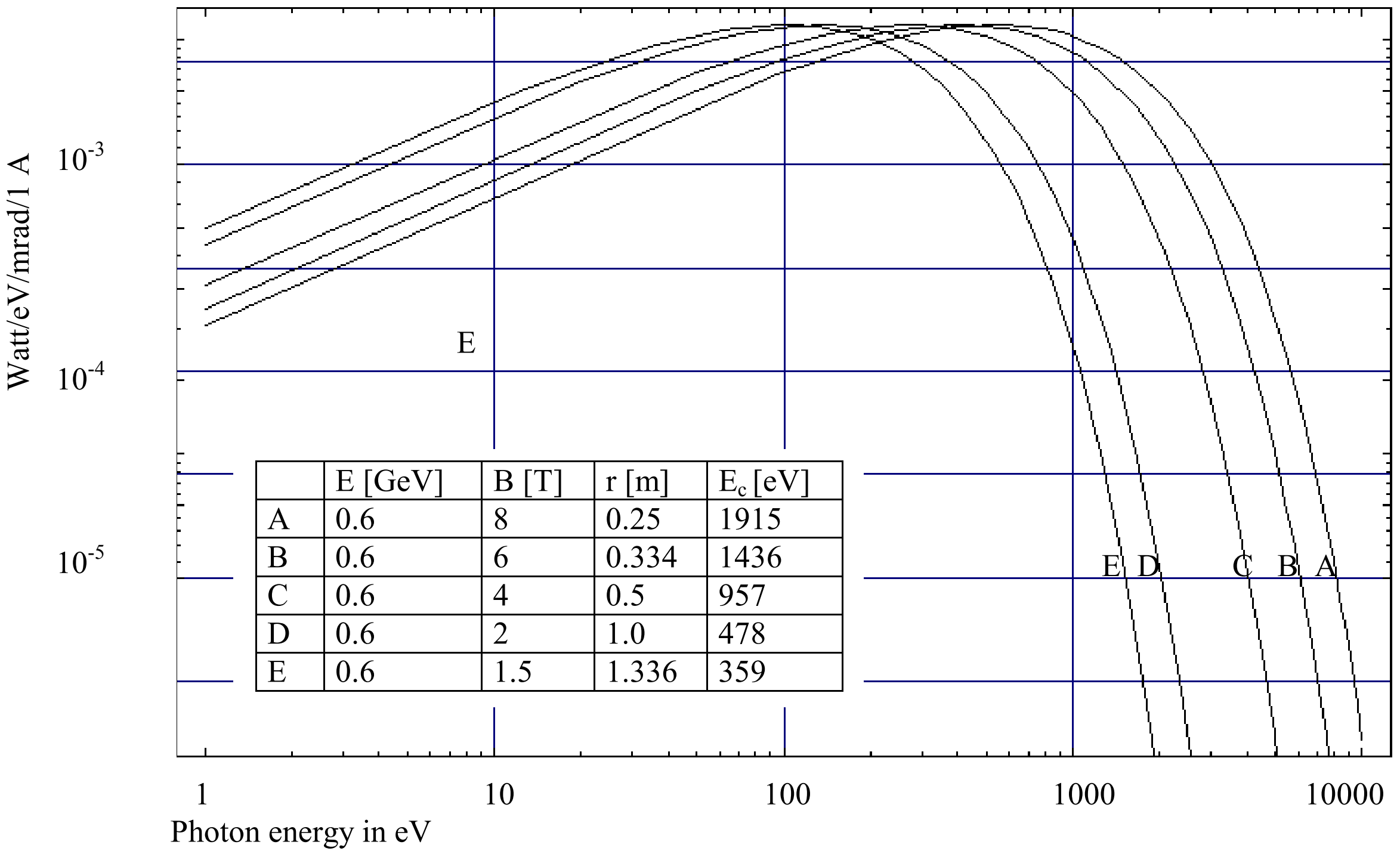

| | E [GeV] | B [T] | r [m] | $E_c$ [eV] |
|---|---|---|---|---|
| A | 0.6 | 8 | 0.25 | 1915 |
| B | 0.6 | 6 | 0.334 | 1436 |
| C | 0.6 | 4 | 0.5 | 957 |
| D | 0.6 | 2 | 1.0 | 478 |
| E | 0.6 | 1.5 | 1.336 | 359 |

*Fig. 3*

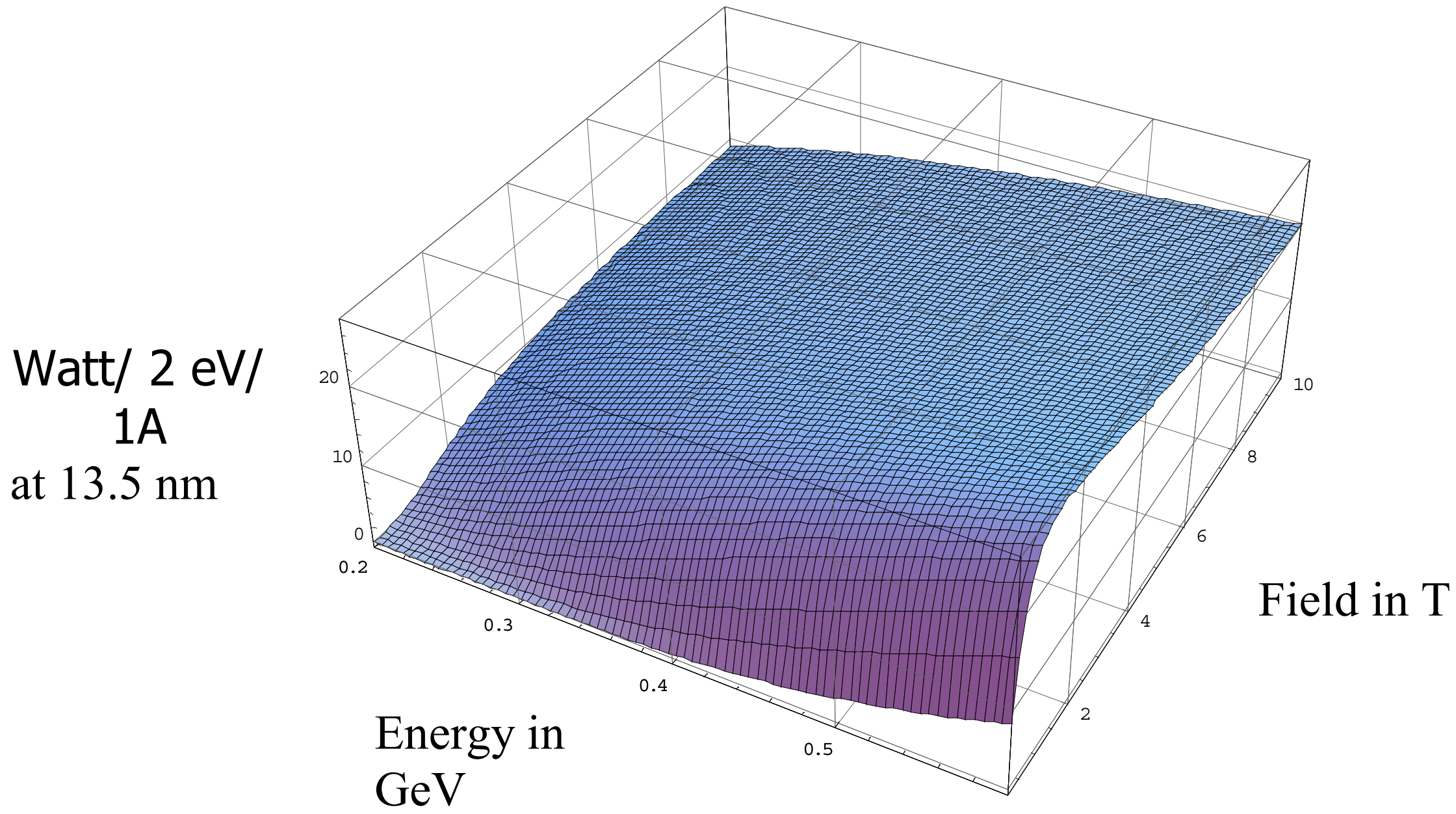

*Fig. 4*

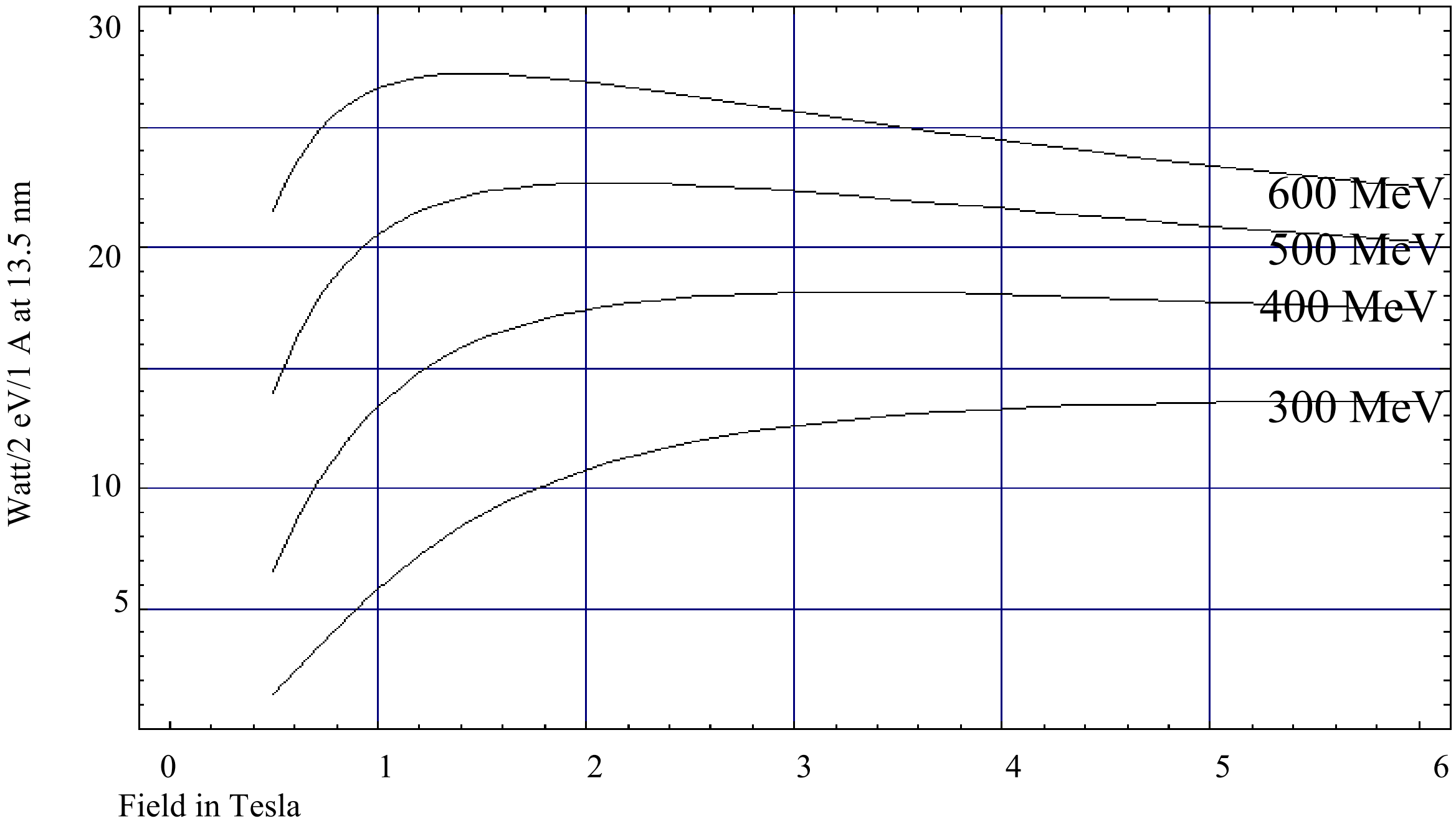

*Fig. 5*

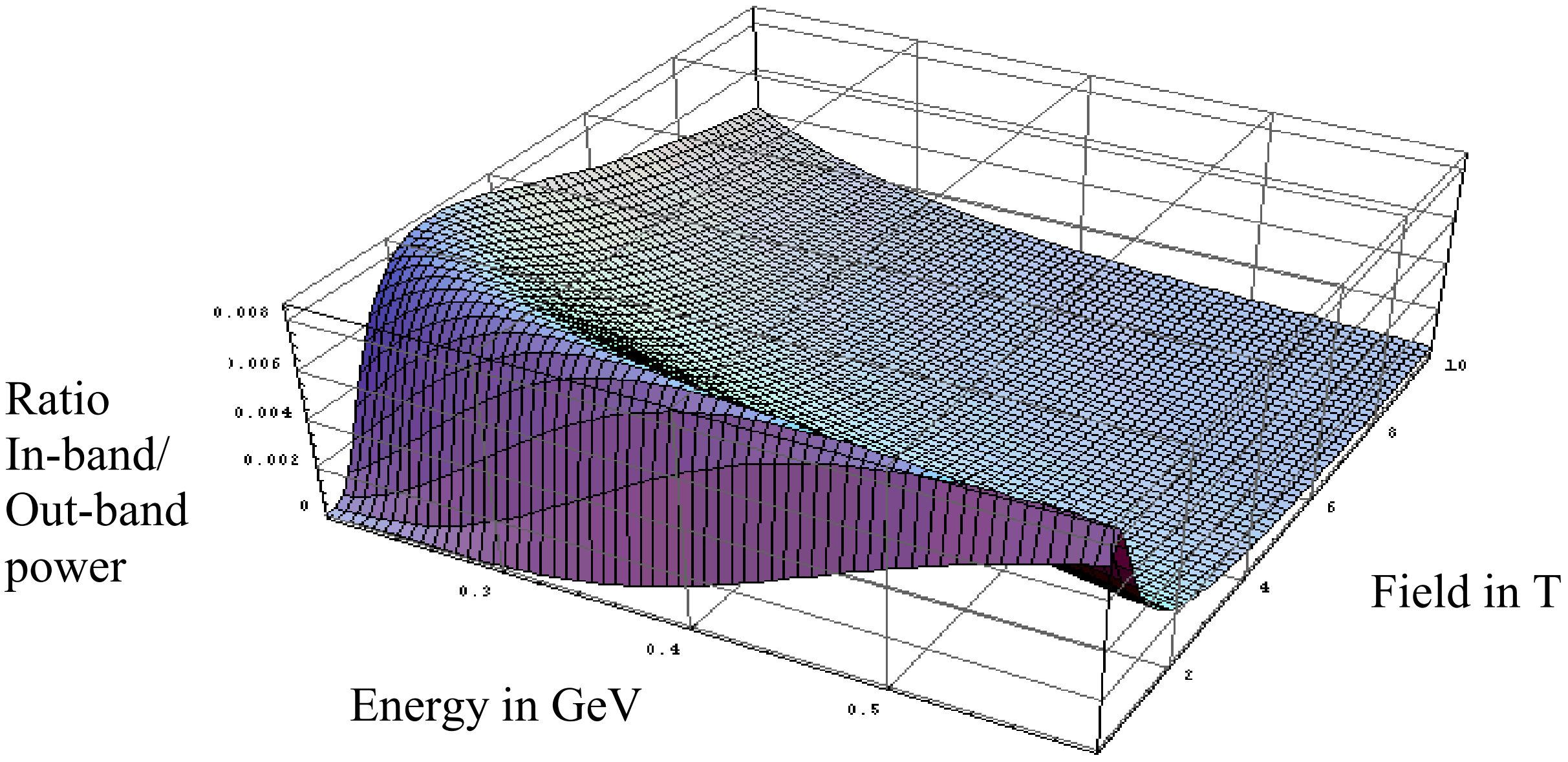

*Fig. 6*

| | |
|---|---|
| Beam energy | 885 MeV |
| Period length $\lambda_u$ | 3.8 mm |
| Number of periods $N_u$ | 100 |
| Field | 0.3 T |
| Gap | 2 mm |

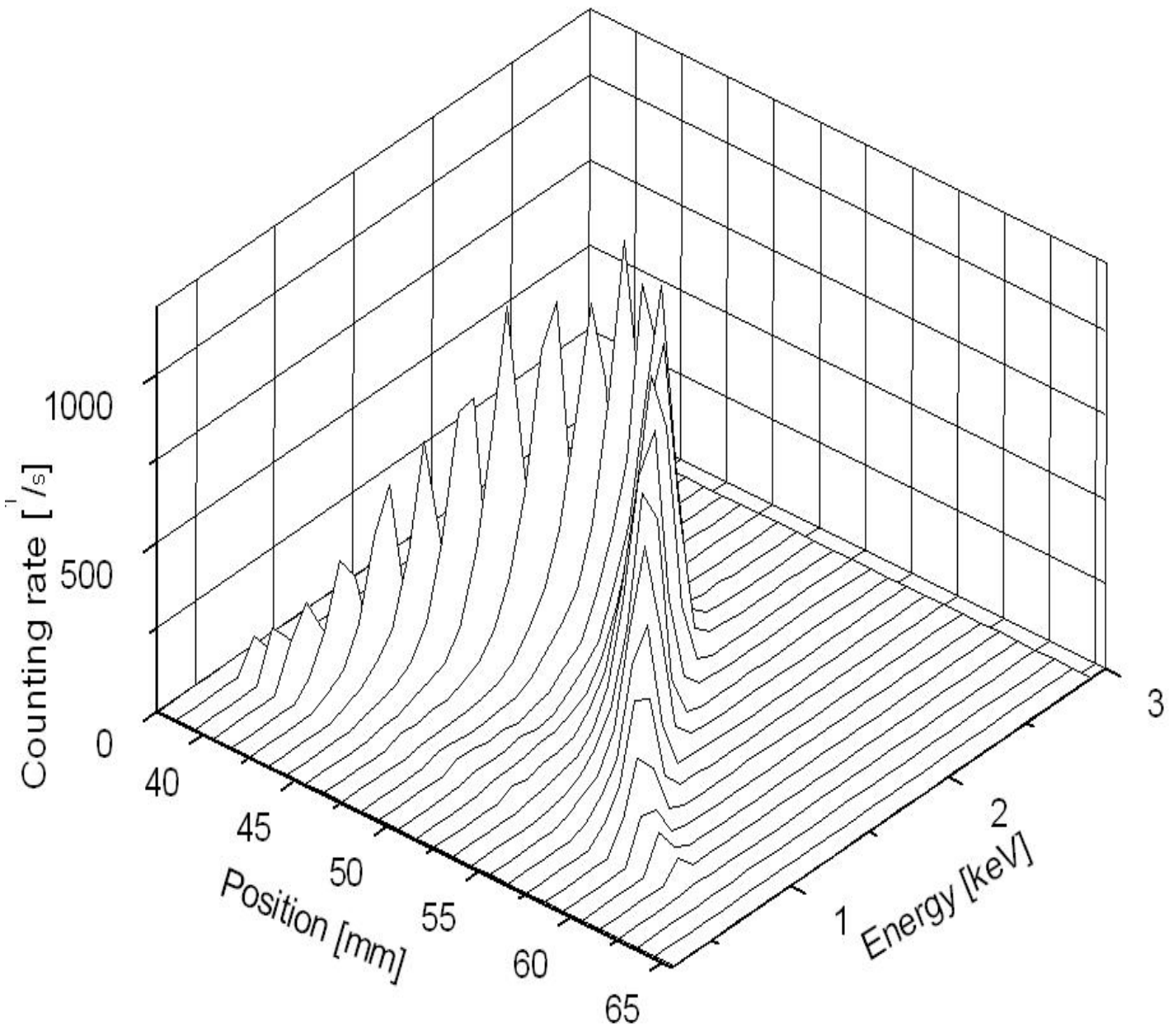

*Fig. 7*

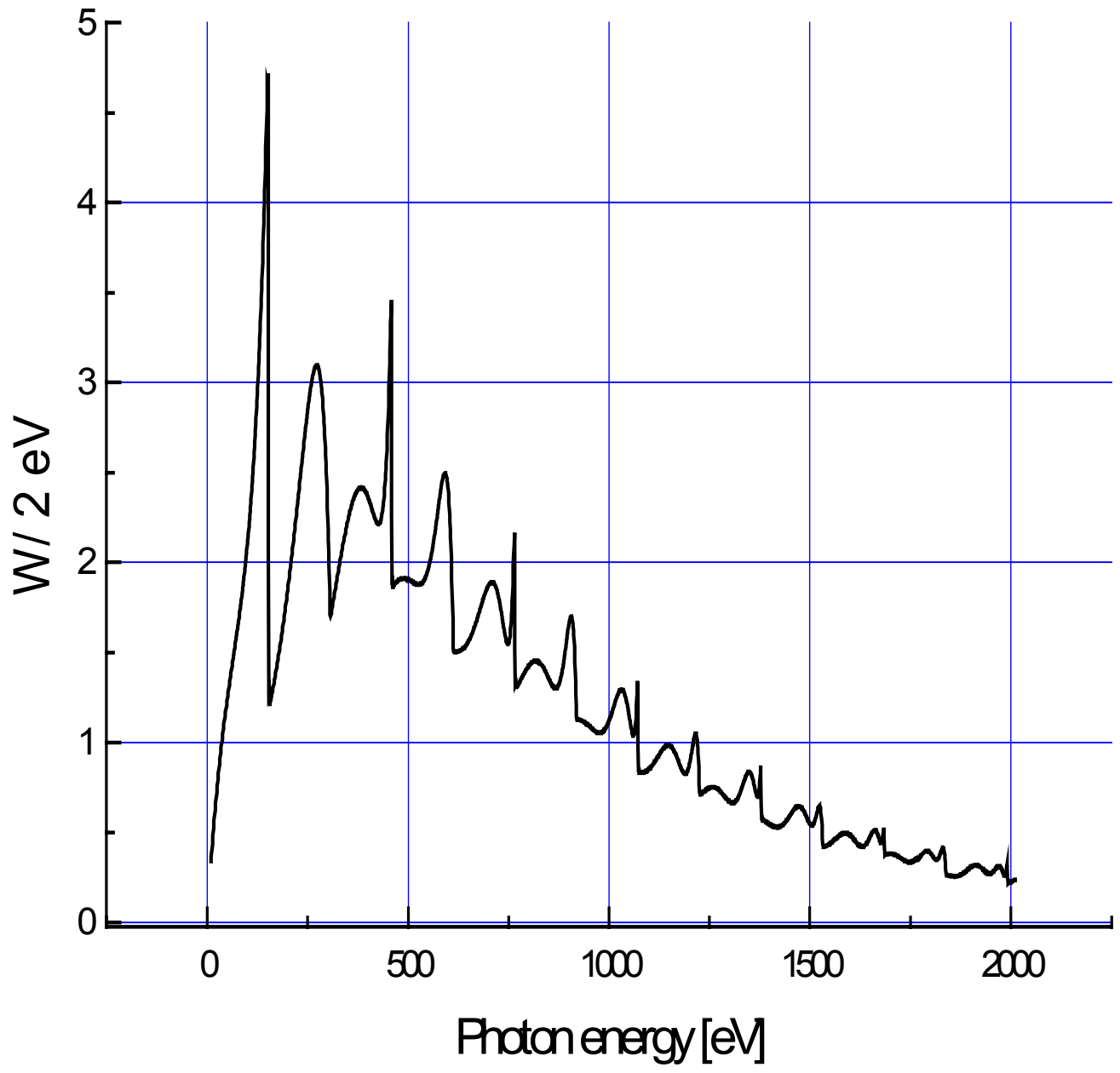

*Fig. 8*

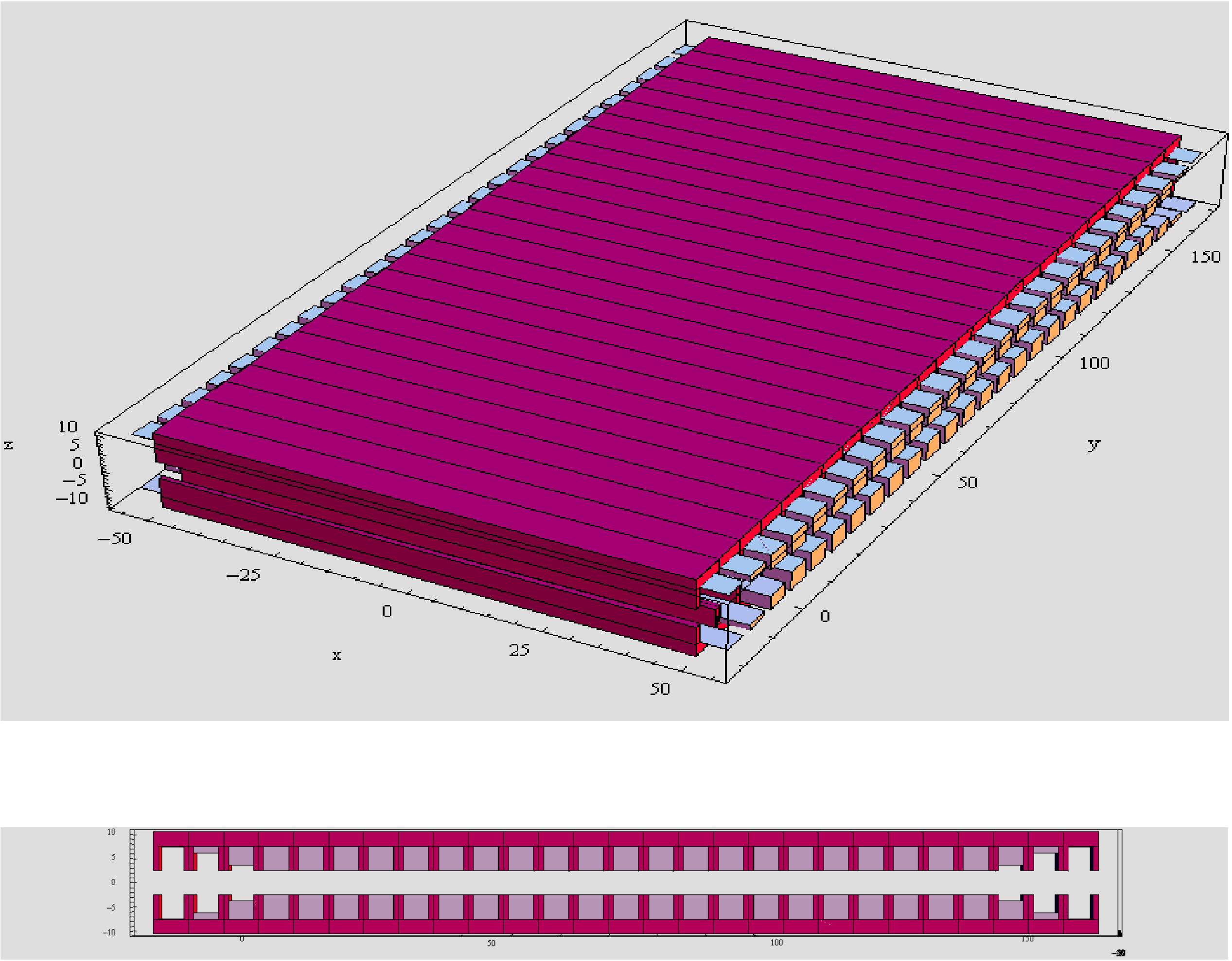

*Fig. 9*

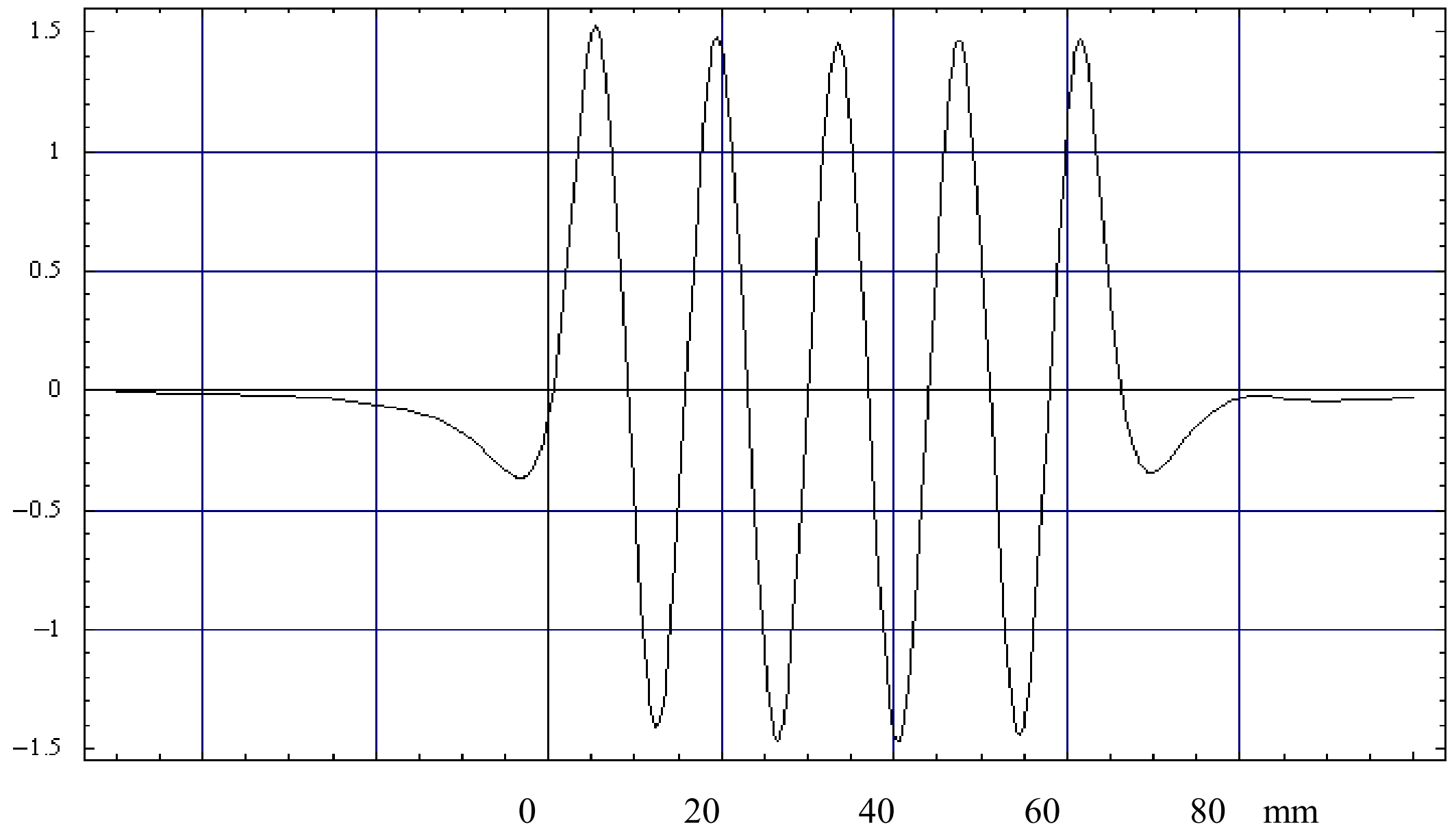

*Fig. 10*

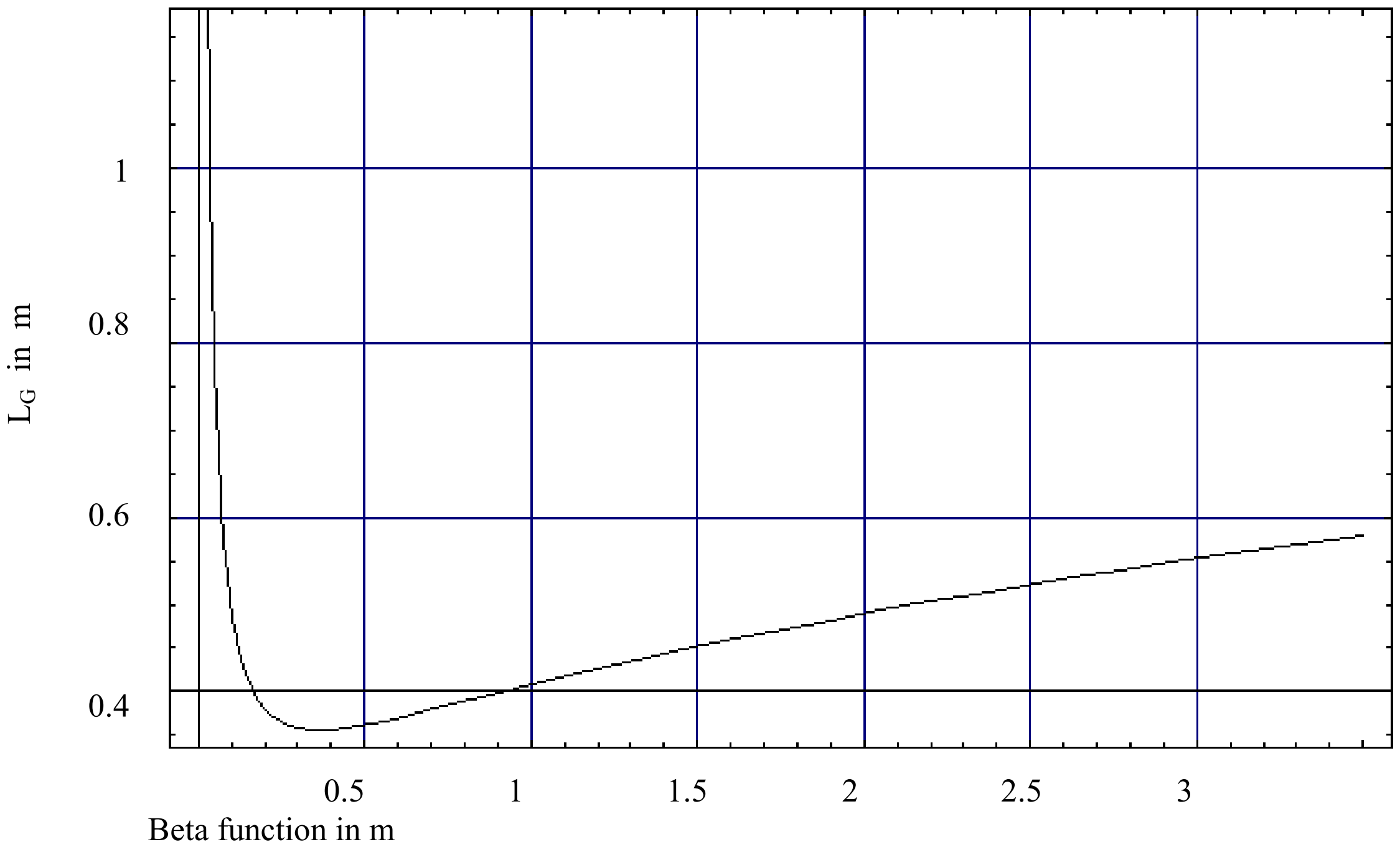

*Fig. 11*

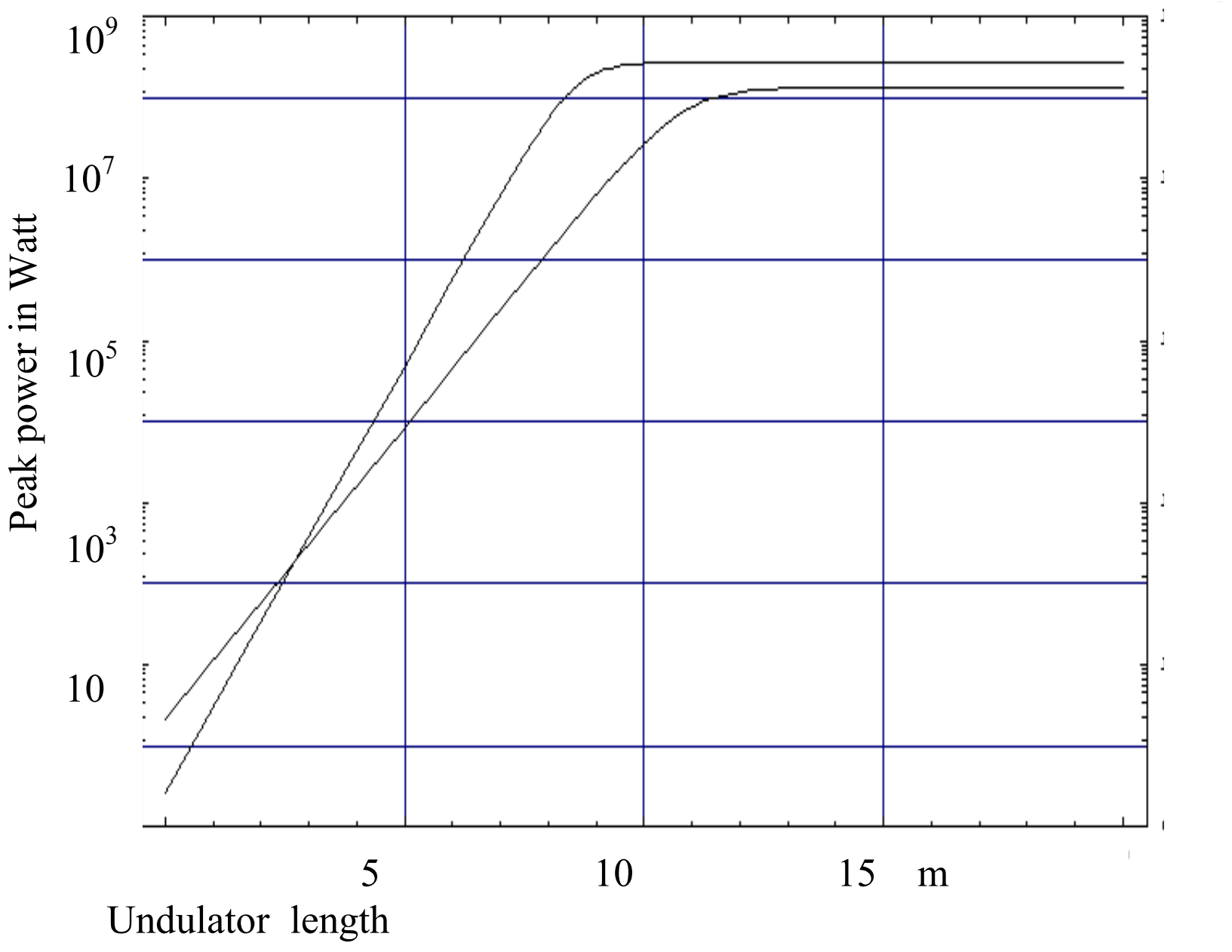

*Fig. 12*

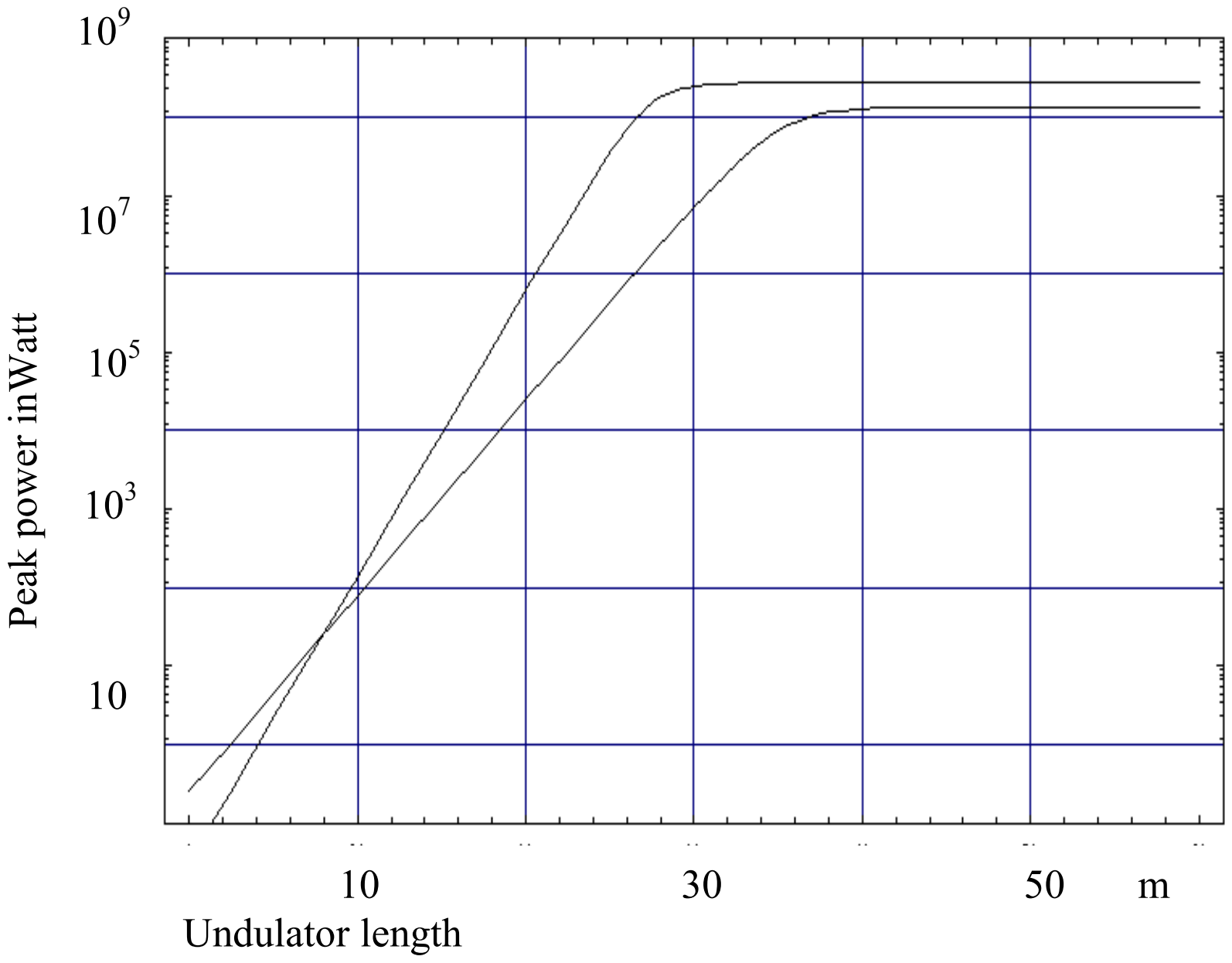

*Fig. 13*

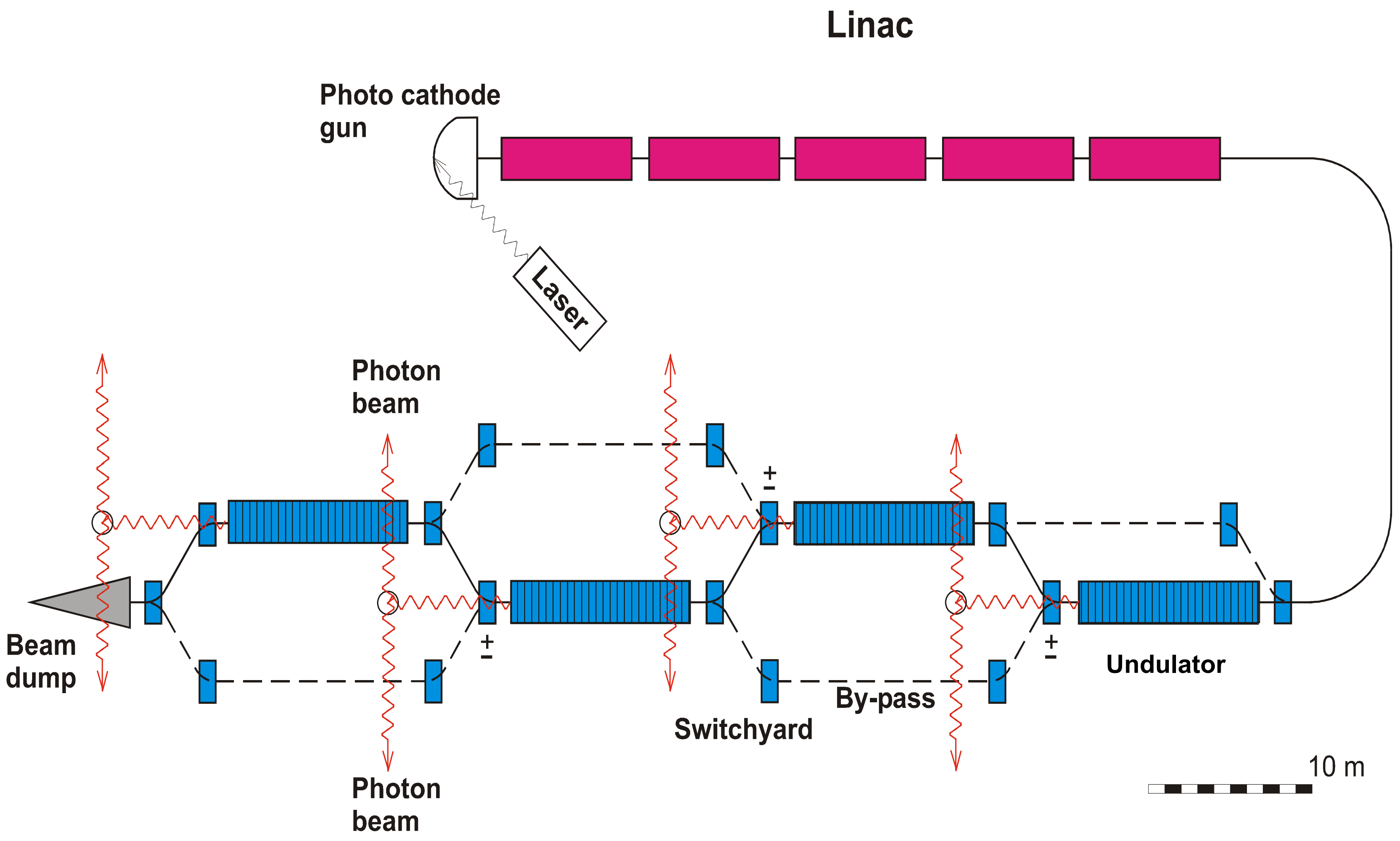

*Fig 14*